\documentclass[aps,pra,twocolumn,groupedaddress,nofootinbib,longbibliography]{revtex4-1}
\usepackage{amsmath}
\usepackage{amssymb}
\usepackage{graphicx}
\usepackage{mathrsfs}
\usepackage{ntheorem}
\usepackage{mathtools}
\usepackage{enumerate}
\usepackage{enumitem}
\usepackage[T1]{fontenc}
\usepackage{physics}
\usepackage{subfigure}
\usepackage{overpic}
\usepackage{epstopdf}
\usepackage{bm}
\usepackage{color,soul}
\usepackage[bookmarks=false]{hyperref}
\usepackage{xcolor}
\hypersetup{colorlinks=true,citecolor=blue,
linkcolor=blue,urlcolor=blue,pdfstartview=FitH,
bookmarksopen=true}

\begin{document}
\title{Basis-independent quantum coherence and its distribution under relativistic motion}
\author{Ming-Ming Du$^1$}
\email{mingmingdu@njupt.edu.cn}
\author{Hong-Wei Li$^1$, Zhen Tao$^1$, Shu-Ting Shen$^1$, Xiao-Jing Yan$^2$, Xi-Yun Li$^2$, Wei Zhong$^3$, Yu-Bo Sheng$^{1,3}$}
\author{Lan Zhou$^2$}
\email{zhoul@njupt.edu.cn}

\affiliation{$1.$ College of Electronic and Optical Engineering and College of Flexible Electronics (Future Technology), Nanjing
University of Posts and Telecommunications, Nanjing, 210023, China\\
$2.$ School of Science, Nanjing University of Posts and Telecommunications, Nanjing,
210023, China\\
$3.$ Institute of Quantum Information and Technology, Nanjing University of Posts and Telecommunications, Nanjing, 210003, China}
\date{\today}
\begin{abstract}
Recent studies have increasingly focused on the effect of relativistic motion on quantum coherence. Prior research predominantly examined the influence of relative motion on basis-dependent quantum coherence, underscoring its susceptibility to decoherence under accelerated conditions. Yet, the effect of relativistic motion on basis-independent quantum coherence, which is critical for understanding the intrinsic quantum features of a system, remains an interesting open question.
This paper addresses this question by examining how total, collective, and localized coherence are affected by acceleration and coupling strength.
 Our analysis reveals that both total and collective coherence significantly decrease with increasing acceleration and coupling strength, ultimately vanishing at high levels of acceleration. This underscores the profound impact of Unruh thermal noise.
 Conversely, localized coherence exhibits relative stability, decreasing to zero only under the extreme condition of infinite acceleration. Moreover, we demonstrate that collective, localized, and basis-independent coherence collectively satisfy the triangle inequality. These findings are crucial for enhancing our understanding of quantum information dynamics in environments subjected to high acceleration and offer valuable insights on the behavior of quantum coherence under relativistic conditions.

\end{abstract}

\maketitle

\section{Introduction}
The study of quantum coherence within relativistic frameworks represents an intriguing frontier in modern physics, integrating elements of quantum information theory, general relativity, and quantum field theory.
Quantum coherence quantifies a state's ability to exhibit quantum interference phenomena and is foundational for developing nanoscale thermodynamics\cite{Lostaglio2015}, quantum metrology\cite{Giovannetti2004}, and quantum biology\cite{Lloyd2011,Huelga2013}.
Traditionally, studies have focused on quantum coherence under non-relativistic, inertial conditions\cite{Streltsov2017,Chitambar2019}.
However, as quantum technologies evolve toward regimes where relativistic effects are significant, comprehending coherence in these contexts is crucial.

The Unruh effect\cite{Unruh1984}, a phenomenon predicted by quantum field theory in curved spacetime, states that an accelerating observer will perceive the vacuum as a thermal bath of particles. This effect introduces novel decoherence mechanisms that can degrade quantum coherence, thereby impacting the stability and performance of quantum systems in relativistic motion. Extensive prior research has documented the effects of acceleration on entanglement\cite{Bruschi1,Bruschi2,FuentesSchuller2005,Alsing2006,Adesso2007,Mann2006,Pan2008,Pan2008a,MartinMartinez2009,Wang2009,
Fuentes2010,Friis2011,MartinMartinez2011,Wang2011,Hwang2011,Chang2012,Tian2012,Xu2014,Richter2015,Regula2016,Richter2017,
He2018,Wu2023,Barman2023,Bak2024,Sen2024,Wu2024,Li2018,TorresArenas2019,Wu2022,Wu2022a,Liu2022,Wu2022b}, quantum discord\cite{Datta2009,Wang2010,Wang2014,Qiang2015}, quantum steering\cite{Wu2022c,Hu2021,Wang2017,Liu2018,Sun2017,Wang2016}, Bell nonlocality\cite{Friis2011,Tian2012,Wu2022b,Dunningham2009,Smith2012,Tian2013,He2016,Zhang2023,He2024,Wang2018} and basis-dependent coherence\cite{Lostaglio2015,Lloyd2011,Streltsov2017,He2024,Wang2016a,Wang2018,Wu2019,Fan2019,Bao2021,Liu2021,Xiao2022,AbdRabbou2022,Harikrishnan2022,Wu2023a,He2018a,Wu2019a,Wu2020,Wu2021,Zeng2021,Wu2022d,Wu2023b}. These studies reveal that quantum correlations and basis-dependent coherence are degraded as relative acceleration increases.

Despite these advancements, the exploration of basis-independent quantum coherence dynamics under relativistic conditions remains limited. Basis-independent coherence\cite{Designolle2021,Radhakrishnan2019,Yao2015,Ma2019,Radhakrishnan2016} is a universal property and does not depend on the choice of basis. It is  crucial for understanding the inherent quantum features of a system, like in foundational quantum physics\cite{Radhakrishnan2019,Yin2022,Yin2021,Singh2021,Yao2015,Ma2019}.

In this paper, we fill the gap by examining how different forms of quantum coherence—total, collective, and localized—are affected by changes in acceleration and the coupling strength within quantum systems. Our analysis utilizes the theoretical framework of Unruh-DeWitt detectors, a model extensively employed to study interactions between quantum fields and accelerating observers. We investigate how Unruh thermal noise, induced by acceleration, affects these coherence measures. Significantly, our findings reveal a profound degradation of both total and collective coherence with increasing acceleration and coupling strength, with these coherence types vanishing entirely at high levels of acceleration.Conversely, localized coherence demonstrates remarkable stability, declining to zero solely in the extreme condition of infinite acceleration. Furthermore, we provide the demonstration that collective, localized, and basis-independent coherence collectively satisfy the triangle inequality. These results provide essential insights into the behavior of quantum coherence under relativistic conditions and enhance our understanding of quantum information dynamics in high-acceleration environments.

To be self-contained, we organize the rest of this paper as follows. In Section \ref{sec1}, we provide a concise overview of coherence measures and explore the quantum information framework of entangled Unruh-DeWitt detectors, as well as the evolution of a prepared state when only one detector is in relativistic motion. In Sec. \ref{sec3}, we investigate the dynamics of total quantum coherence, collective coherence, and localized coherence under Unruh thermal noise. Finally, we conclude this work in Sec. \ref{sec4}.
\section{PRELIMINARIES}\label{sec1}
\subsection{The measures of total, collective, and localized coherence}
To present our finding clearly, we first review the measures of total, collective, and localized coherence. Quantifying coherence within a quantum state provides profound insights into the state's underlying quantum properties. One of the fundamental measures is the total quantum coherence, denoted as $ \mathrm{C}_\mathrm{T}(\rho) $, which is significant for its methodological implications in quantum information theory. This measure is defined as\cite{Radhakrishnan2019,Radhakrishnan2016}:
\begin{align}\label{du4}
\mathrm{C}_\mathrm{T}\left(\rho\right) = \sqrt{S\left(\frac{\rho + \rho_I}{2}\right) - \frac{\mathrm{S}\left(\rho\right) + \log_2 d}{2}},
\end{align}
where $S(\cdot) $ represents the von Neumann entropy, $ \rho_I= I/d$ is the maximally mixed state in a $d$-dimensional Hilbert space, and $\log_2d$ is the logarithmic base-$2$ of $ d $. This formula measures the "distance" in entropy between the quantum state $\rho$ and the maximally mixed state $\rho_I $, thereby quantifying the total coherence regardless of the chosen basis.

Expanding the concept of coherence to multipartite systems, where a state comprises several subsystems, introduces collective coherence\cite{Radhakrishnan2019,Radhakrishnan2016}, denoted as $C_C(\rho)$. This is defined by:
\begin{align}\label{du5}
C_C(\rho) = \sqrt{S\left(\frac{\rho + \pi_\rho}{2}\right) - \frac{S(\rho) + S(\pi_\rho)}{2}},
\end{align}
where $\pi_\rho = \rho_I \otimes \rho_2 \ldots \otimes \rho_n$ is the tensor product of the reduced density matrices $\rho_i$ for each subsystem. This expression evaluates the coherence due to the interaction and correlation among the subsystems, which goes beyond what is contributed individually by the subsystems. The measure $ C_C(\rho) $ thus serves as a crucial tool for understanding the entangled and collective quantum effects that are not apparent when subsystems are considered in isolation.

Parallelly, localized coherence\cite{Radhakrishnan2019,Radhakrishnan2016}, $ C_L(\rho) $, focuses on the coherence within individual subsystems isolated from their collective dynamics. It is expressed as:
\begin{align}\label{du6}
C_{L}\left(\rho\right) = \sqrt{S\left(\frac{\pi_\rho + \rho_I}{2}\right) - \frac{S(\pi_\rho) + \log_2 d}{2}},
\end{align}
This formula contrasts the closest product state $\pi_\rho$ with the maximally mixed state $\rho_I$, thereby isolating the coherence attributable to intrinsic subsystem properties rather than emergent inter-subsystem interactions.

\subsection{Modeling the Evolution of Quantum Detectors under Relativistic motion}\label{sec2}
We consider a scenario involving two observers, Alice and Bob, each equipped with an Unruh-DeWitt detector\cite{Unruh1984} modeled through a two-level noninteracting atom\cite{landulfo2009,celeri2010,Wang2016a,Tian2013}, in Minkowski space-time. Alice's detector remains stationary, while Bob's is subject to uniform acceleration $a$ along the $x$ axis for a duration $\Delta$. We then have Alice's detector always off and Bob's detector on when it moves with constant acceleration.
The worldline of Bob's detector, described in Rindler spacetime coordinates, is given by: $ t(\tau)= a^{-1} \sinh(a\tau), x(\tau)=a^{-1}\cosh(a\tau), y(\tau)= z(\tau) = 0 $, where  $\tau$ is the qubit proper time and $(t,x,y,z)$  are the usual Cartesian coordinates of Minkowski space-time. Rindler spacetime describes the perspective of a uniformly accelerating observer and contrasts with Minkowski spacetime, which represents an inertial (non-accelerating) observer's view.
The initial state of the detector-field system is assumed to be:
\begin{align}\label{du3}
\left|\Psi_{t_0}^{AR\phi}\right\rangle = \left|\Psi_{AR}\right\rangle \otimes \left|0_M\right\rangle,
\end{align}
where $|\Psi_{AR}\rangle = 1/\sqrt{2} (|0_A\rangle |1_R\rangle +|1_A\rangle|0_R\rangle)$ denotes the entangled initial state shared between Alice's (A) and Bob's (R) detectors, and $|0_M\rangle$ represents the external scalar ﬁeld is in Minkowski vacuum.

The total Hamiltonian for the system, denoted as $ H_{AR\phi} $, is comprised of several components\cite{landulfo2009,celeri2010,Wang2016a,Tian2013}:
\begin{align}\label{du1}
H_{A\:R\:\phi} = H_A + H_R + H_{KG} + H_{\mathrm{int}}^{R\phi},
\end{align}
where $ H_A = \Omega A^\dagger A $ and $ H_R = \Omega R^\dagger R $ represent the Hamiltonians for the detectors. Here, $\Omega$ is the energy gap of the detectors.
 The interaction Hamiltonian \( H_{int}^{R\phi}(t) \) for the accelerated detector, moving in Rindler spacetime, is
   \begin{align}
   H_{int}^{R\phi}(t) = \epsilon(t) \int_{\Sigma_t} d^3\mathbf{x} \sqrt{-g} \phi(x) [\chi(\mathbf{x}) R + \overline{\chi}(\mathbf{x}) R^\dagger],
   \end{align}
 where $\epsilon(t)$ is the coupling strength, $\phi(x)$ is the free Klein-Gordon field operator, and $\chi(\mathbf{x})$ a Gaussian profile localizing the interaction. This Hamiltonian captures how an accelerating detector interacts with the scalar quantum field, emphasizing differences in how quantum information is processed in non-inertial frames compared to inertial frames in Minkowski spacetime. Such a Gaussian coupling function describes a pointlike detector which only interacts with the neighbor scalar ﬁelds in the Minkowski vacuum.

In the case of weak coupling, we can calculate the final state$|\Psi_{t=t_0+\Delta}^{R\phi}\rangle$ of the atom-field system at time $t=t_0+\Delta $  in the ﬁrst order of perturbation over the coupling constant $ \epsilon$. Under the dynamic evolution described by the Hamiltonian given by Eq. (\ref{du1}), the final state   $|\Psi_t^{R\phi}\rangle$ can be obtained as:
\begin{align}\label{du2}
\left|\Psi_t^{R\phi}\right\rangle = \left\{ I - i\left[\phi(f) R + \phi(f)^\dagger R^\dagger\right] \right\} \left|\Psi_{t_0}^{R\phi}\right\rangle,
\end{align}
where $\phi(f)$ is the field operator, which is defined as
\begin{align}
\phi(f) \equiv \int d^4x \sqrt{-g} \phi(x) f = i \left[a_{RI}(\overline{u E \overline{f}}) - a_{RI}^\dagger(u E f)\right],
\end{align}
described in Eq. (\ref{du2}). This operator characterizes the distribution of the external scalar field in the system. Here, $a_{RI}(\overline{u}) $ and $ a_{RI}^\dagger(u) $ are the annihilation and creation operators of the $ u $-modes, respectively, and $ f \equiv \epsilon(t)e^{-i\Omega t}\chi(\mathbf{x}) $ is a compactly supported complex function defined in Minkowski space-time. Furthermore, $u$ represents the operator that isolates the positive-frequency components of the solutions to the Klein-Gordon equation in Rindler metric\cite{landulfo2009,R}, and $E$ denotes the difference between the advanced and retarded Green's functions.

Upon substituting the initial state from Eq. (\ref{du3}) into Eq. (\ref{du2}), the resultant expression for the final state of the total system, leveraging Rindler operators \( a_{RI}^\dagger \) and \( a_{RI} \), is articulated as:
\begin{align}
\left|\Psi_{t}^{AR\phi}\right\rangle &= \left|\Psi_{t_0}^{AR\phi}\right\rangle + \frac{1}{\sqrt{2}} \left|0_A\right\rangle \left|0_R\right\rangle \otimes \left[a_{RI}^\dagger(\lambda) \left|0_M\right\rangle\right] \\\notag
&+ \frac{1}{\sqrt{2}}\left|1_A\right\rangle \left|1_R\right\rangle \otimes \left[a_{RI}(\overline{\lambda}) \left|0_M\right\rangle\right],
\end{align}
where $\lambda = -uEf$. Here, the Rindler operators $a_{RI}^\dagger(\lambda)$ and $a_{RI}(\overline{\lambda})$ are operational within Rindler region $I$, and $\left|0_M\right\rangle$ denotes the vacuum state in the Minkowski space-time. The Bogoliubov transformations linking the Rindler and Minkowski operators are given by\cite{landulfo2009,celeri2010,Wang2016a,Tian2013}:
\begin{align}
a_{RI}(\overline{\lambda}) &= \frac{a_M(\overline{F_{1\Omega}}) + e^{-\pi\Omega/a} a_M^\dagger(F_{2\Omega})}{(1-e^{-2\pi\Omega/a})^{1/2}},\\
a_{RI}^\dagger(\lambda) &= \frac{a_M^\dagger(F_{1\Omega}) + e^{-\pi\Omega/a} a_M(\overline{F_{2\Omega}})}{(1-e^{-2\pi\Omega/a})^{1/2}},
\end{align}
where
\begin{align}
F_{1\Omega} &= \frac{\lambda + e^{-\pi\Omega/a} \lambda \circ w}{(1-e^{-2\pi\Omega/a})^{1/2}}, \\
F_{2\Omega} &= \frac{\overline{\lambda \circ w} + e^{-\pi\Omega/a} \overline{\lambda}}{(1-e^{-2\pi\Omega/a})^{1/2}}.
\end{align}
Here, $w(t,x) = (-t, -x)$ is a wedge reflection isometry, which reflects the function $\lambda$ from Rindler region $I$ to $\lambda \circ w$ in Rindler region $II$\cite{landulfo2009,celeri2010,Wang2016a,Tian2013}.

We focus on the dynamics of the detectors' states post interaction with the external field. By tracing out the external field \(\phi(f)\)'s degrees of freedom, we derive the density matrix representing the state of the detectors,

\begin{align}
\rho_{t}^{AR} = \begin{pmatrix}
\gamma & 0 & 0 & 0 \\
0 & \alpha & \alpha & 0 \\
0 & \alpha & \alpha & 0 \\
0 & 0 & 0 & \beta
\end{pmatrix},
\end{align}
where $\Psi_{AR}$ denotes the initial state of the detectors as specified in Eq. (\ref{du3}). The parameters $\alpha$, $\beta$, and $\gamma$ are defined as follows:
\begin{align}
\alpha &= \frac{1-q}{2(1-q) + \nu^2(1 + q)},\\\notag
\beta &= \frac{\nu^2q}{2(1-q) + \nu^2(1 + q)},\\\notag
\gamma&=\frac{\nu^2}{2(1-q) + \nu^2(1 + q)},
\end{align}
where the parametrized acceleration $q$ is expressed as $q \equiv e^{-2\pi\Omega/a}$, and the effective coupling $\nu^2$ is given by $\nu^2 \equiv \|\lambda\|^2 = \epsilon^2\Omega\Delta e^{-\Omega^2\kappa^2}/(2\pi)$\cite{landulfo2009,celeri2010,Wang2016a,Tian2013,R}, with the condition $\Omega^{-1} \ll \Delta$ necessary for the validity of this definition. Furthermore, the effective coupling should be restricted to $\nu^2 \ll 1$ for the validity of the perturbative approach applied in this paper. Notably, $q$ is a monotonic function of the acceleration parameter $a$; specifically, $q$ approaches $0$ in the limit of zero acceleration and tends towards $1$ as the acceleration becomes infinitely large.

\section{Impact of Unruh thermal Noise on Quantum Coherence Measures}\label{sec3}
We aim to investigate the dynamics of total quantum coherence ($C_T$), collective coherence ($C_C$), and localized coherence ($C_L$) under Unruh thermal noise, analyzing their interrelationships. Using Eqs. (\ref{du4}-\ref{du6}), we derive these coherence measures:
\begin{align}
C_T(\rho_{t}^{AR}) &= \sqrt{S(\eta_i) - \frac{1}{2}S(\lambda_i) - 1}, \\
C_C(\rho_{t}^{AR}) &= \sqrt{S(\zeta_i) - \frac{1}{2}S(\lambda_i) - \frac{1}{2}S(\vartheta_i)}, \\
C_L(\rho_{t}^{AR}) &= \sqrt{S(\xi_i) - \frac{1}{2}S(\vartheta_i) - 1},
\end{align}

where the eigenvalues are defined as follows: $\eta= \left\{1/8, (1+8\alpha)/8, (1+4\beta)/8, (1+4\gamma)/8\right\}$,
$\lambda = \{0, 2\alpha, \beta, \gamma\}$,
$\zeta = \{\beta+(\alpha+\beta)^2, (\alpha+\beta)(\alpha+\gamma)/2, [\alpha(2+\alpha+\beta)+(\alpha+\beta)\gamma]/2$, $[\gamma+(\alpha+\gamma)^2]/2\}$,
$\vartheta = \{\alpha^2, (\alpha+\beta)(\alpha+\gamma), (\alpha+\beta)(\alpha+\gamma)$,$(\alpha+\gamma)^2\}$,
$\xi = \{[1+4(\alpha+\beta)^2]/8, [1+4(\alpha+\beta)(\alpha+\gamma)]/8, [1+4(\alpha+\beta)(\alpha+\gamma)]/8, [1+4(\alpha+\gamma)^2]/8\}$.

\begin{figure}
\includegraphics[width=0.44\textwidth]{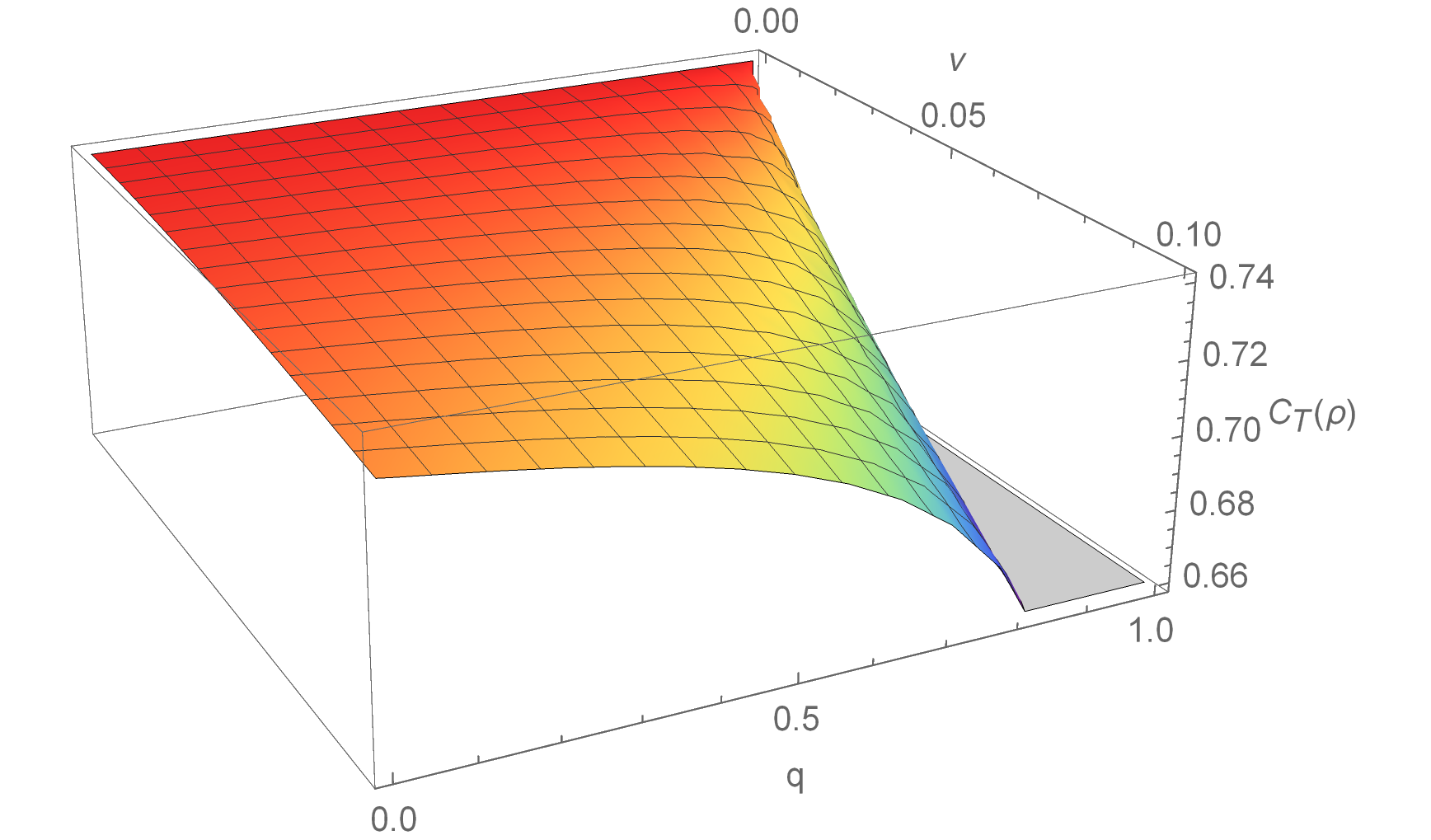}
\caption{ Total quantum coherence between the detectors as functions of the effective coupling parameter $\nu$ and the acceleration parameter $q$.}
\label{fig1}
\end{figure}

Total Quantum Coherence ($C_T$) reflects the overall coherence of the quantum state across the entire system, highlighting how quantum superpositions are maintained or decay. In Fig.~(\ref{fig1}), we illustrate how $C_T$ between the detectors varies as a function of the effective coupling parameter $\nu$ and the acceleration parameter $q$. Initially, observe that at zero acceleration with non-zero effective coupling $\nu$, total coherence deviates from $\sqrt{1/8-5\log{(5/8)}/8}$, anticipated for the singlet state (\ref{du3}). This is so because even inertial detectors have a nonzero probability of spontaneously decaying (along the nonzero time interval $\Delta$) with the emission of a Minkowski particle, which carries away some information. Such a process leads to a purity loss of the initial singlet state degradating the initial coherence. It is crucial to note that the standard coherence value $\sqrt{1/8-5\log{(5/8)}/8}$ for the singlet state is achieved when $\nu$ is zero, as shown in Fig. (\ref{fig1}). In such a case there is no interaction between Bob’s qubit and the scalar field. From Fig. ~(\ref{fig1}), we can see that $C_T$ monotonically decreases with increases in both $\nu$ and $q$. Importantly, the total quantum coherence reduces to zero at a finite acceleration, indicating a complete loss of coherence due to significant thermal noise induced by acceleration. From the perspective of uniformly accelerated observers, total quantum coherence loss results from Bob's qubit interacting with the Unruh thermal bath of Rindler particles, observed when the field is in the Minkowski vacuum. We recall that the Unruh temperature experienced by the noninertial qubit is proportional to its proper acceleration. Now from the inertial observers’ perspective, the total quantum coherence is carried away by the scalar radiation emitted by the accelerating qubit when it suffers a transition. 
\begin{figure}
\includegraphics[width=0.44\textwidth]{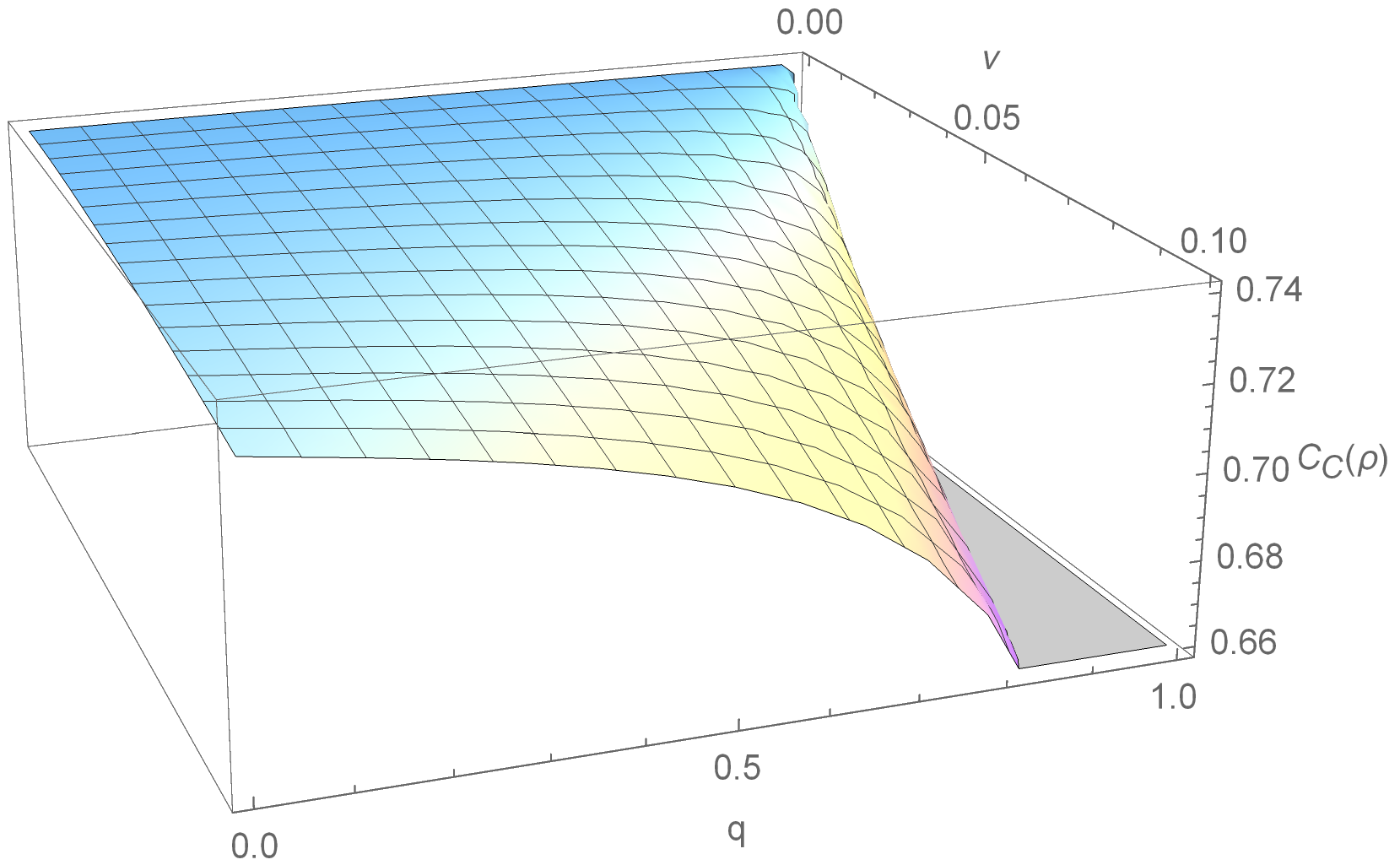}
\caption{Collective coherence between the detectors as functions of the effective coupling parameter $\nu$ and the acceleration parameter $q$.}
\label{fig2}
\end{figure}

 Collective Coherence ($C_C$) focuses on the coherence that arises from correlations between different subsystems, offering insights into how entangled states are influenced by relativistic effects. In Fig.~(\ref{fig2}), we illustrate how $C_T$ between the detectors varies as a function of the effective coupling parameter $\nu$ and the acceleration parameter $q$. we also note that for null acceleration and the effective coupling parameter $\mu\neq0$, the total coherence value differs from $\sqrt{1/8-5\log{(5/8)}/8}$. Moreover, the $C_T$ also decreases with increasing $\nu$ and $q$, paralleling the trend observed for $C_T$. The reasons for these are the similar as total coherence.

\begin{figure}
\includegraphics[width=0.44\textwidth]{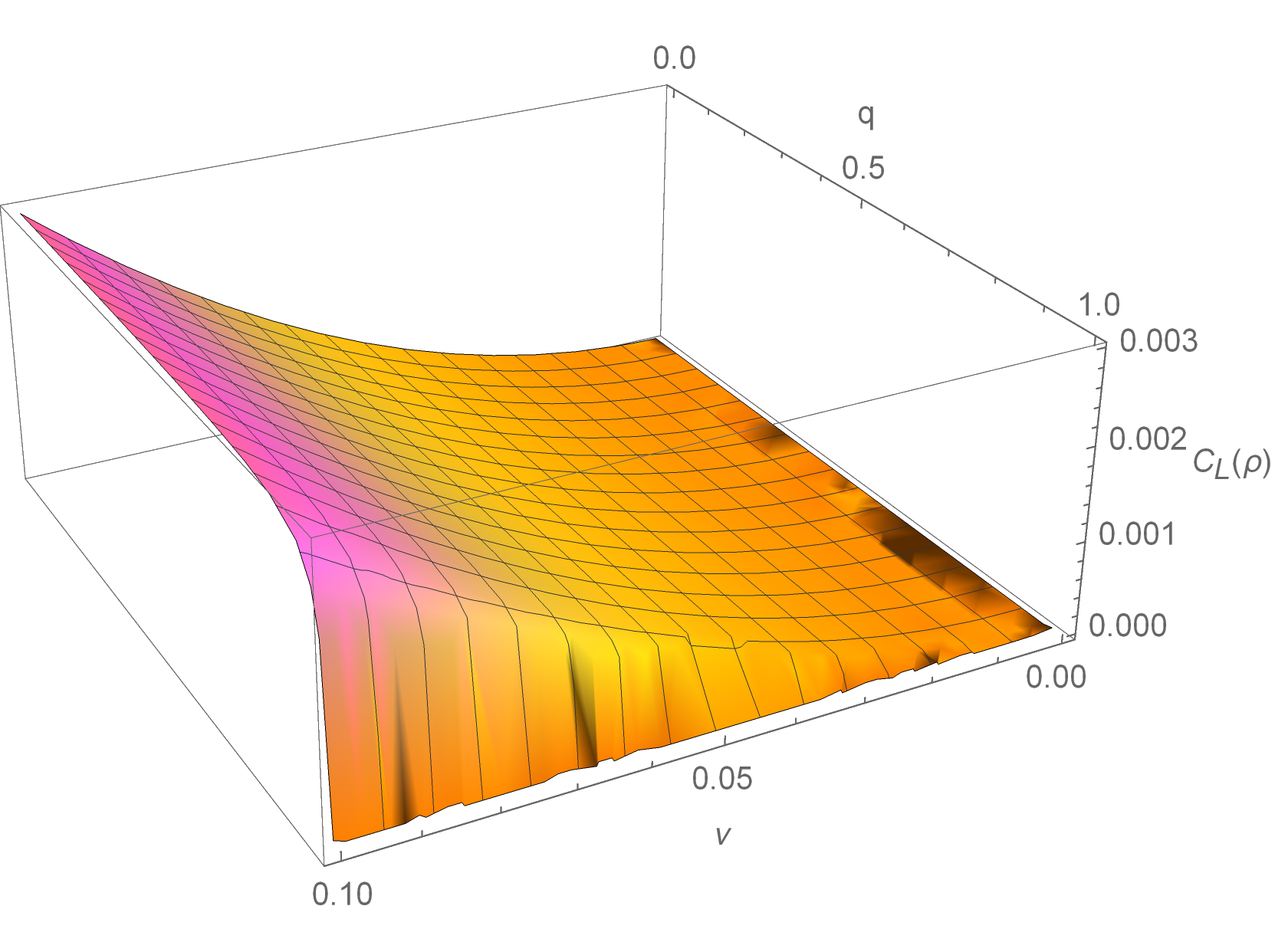}
\caption{Localized coherence between the detectors as functions of the effective coupling parameter $\nu$ and the acceleration parameter $q$.}
\label{fig3}
\end{figure}

Localized Coherence ($C_L$) evaluates the coherence preserved in localized subsystems, indicating the resilience of individual quantum states against external perturbations. Fig.~(\ref{fig3}) details the dynamics of localized coherence ($C_L$) as functions of $\nu$ and $q$. Here, $C_L$ increases with the growth of $\nu$ but decreases as $q$ increases, diverging from the behaviors seen in $C_T$ and $C_C$. Significantly, unlike the other forms of coherence, localized coherence only approaches zero in the limit of infinite acceleration ($q \rightarrow 1$), indicating its comparative robustness against increasing Unruh temperature.

Furthermore, we observe that the total quantum coherence, collective coherence, and localized coherence adhere to the triangle inequality:
\begin{align}
C_C(\rho_{t}^{AR}) + C_L(\rho_{t}^{AR}) \geq C_T(\rho_{t}^{AR}),
\end{align}
underscoring a fundamental geometric relationship between these measures as influenced by quantum dynamical effects under non-inertial conditions. This inequality connects the system's collective and localized behaviors, suggesting that the aggregate coherence of localized subsystems and the whole system sets a lower bound on achievable total coherence.

\section{Conclusions}\label{sec4}
In this study, we conducted a comprehensive investigation into the dynamic behaviors of various quantum coherence measures affected by Unruh thermal noise. Specifically, we analyzed total quantum coherence ($C_T$), collective coherence ($C_C$), and localized coherence ($C_L$), examining their dependency on the effective coupling parameter $\nu$ and the acceleration parameter $q$. Our results show that both $C_T$ and $C_C$ decrease monotonically as $\nu$ and $q$ increase, with $C_T$ reaching zero at a finite acceleration level. This indicates a complete loss of total coherence and collective coherence, underscoring the significant impact of acceleration-induced thermal noise on quantum coherence.

In contrast, localized coherence ($C_L$) increases with $\nu$ but decreases as $q$ approaches unity, uniquely tapering to zero only at this limit. This behavior underscores the robustness of $C_L$ in preserving quantum information within localized subsystems under extreme relativistic conditions, setting it apart from more globally defined coherence measures.

Furthermore, our investigation into the geometric and algebraic interplay among these coherence measures, illustrated through the triangle inequality,$C_C(\rho_{t}^{AR}) + C_L(\rho_{t}^{AR}) \geq C_T(\rho_{t}^{AR})$,
provides profound insights into their structural relationships. This inequality not only enriches the theoretical landscape of quantum coherence but also acts as a practical gauge for assessing the integrity of quantum systems under dynamic conditions.

\begin{acknowledgments}
This work was supported by the National Natural Science Foundation of China(Grant Nos. 12175106 and 92365110), the Natural Science Research Start-up Foundation of Recruiting Talents of Nanjing University of Posts and Telecommunications(Grant No. NY222123), and the Natural Science Foundation of Nanjing University of Posts and Telecommunications(Grant No. NY223069).
\end{acknowledgments}
%
\end{document}